# Against the 'No-Go' Philosophy of Quantum Mechanics

Federico Laudisa

*Department of Human Sciences, University of Milan-Bicocca*

Piazza dell'Ateneo Nuovo 1, 20126 Milan – Italy

federico.laudisa@unimib.it

**Abstract**

In the area of the foundations of quantum mechanics a true industry appears to have developed in the last decades, with the aim of proving as many results as possible concerning what there cannot be in the quantum realm. In principle, the significance of proving 'no-go' results should consist in clarifying the fundamental structure of the theory, by pointing out a class of basic constraints that the theory itself is supposed to satisfy. In the present paper I will discuss some more recent no-go claims and I will argue against the deep significance of these results, with a two-fold strategy. First, I will consider three results concerning respectively local realism, quantum covariance and predictive power in quantum mechanics, and I will try to show how controversial the main conditions of the negative theorem turn out to be – something that strongly undermines the general relevance of these theorems. Second, I will try to discuss what I take to be a common feature of these theoretical enterprises, namely that of aiming at establishing negative results for quantum mechanics in absence of a deeper understanding of the overall ontological content and structure of the theory. I will argue that the only way toward such an understanding may be to cast in advance the problems in a clear and well-defined interpretational framework – which in my view means primarily to specify the ontology that quantum theory is supposed to be about – and after to wonder whether problems that seemed worth pursuing still are so in the framework.

# 1 Introduction

One of the most fascinating aspects of the twentieth-century science concerns the discovery of *impossibilities in principle*. Ranging from Cantor to Planck, from Church to Heisenberg, Gödel and Turing, several are the abstract entities and properties that simply cannot exist or hold in principle, whatever the intellectual resources and the time available to philosophers and scientists might be. As a matter of fact, that dealing with what does *not* exist is a tricky business was already clear at least since the times of the Plato's treatment of non-being, whereas the celebrated Quine's paper "On what there is" (1948) does nothing but recapitulate how difficult it is to defend a non-existence claim, especially when one gets engaged in an ontological dispute. In the area of the foundations of quantum mechanics, however, there seems not to be a similar sensitivity to how controversial a 'negative' result may turn out to be. On the contrary, a true industry appears to have developed in the last decades with the aim of proving as many results as possible concerning what there *cannot* be in the quantum realm. As a matter of fact, there is a rather long sequence of so-called 'no-go' theorems that have been established along the history of quantum mechanics in the last half-century, perhaps partially inspired by the contingent circumstance that the very birth and development of quantum mechanics was based on assumptions and results exhibiting a sort of 'no-go' character (from the Planck discreteness hypothesis concerning the blackbody radiation to the Heisenberg uncertainty relations).

In principle, the significance of proving 'no-go' results should consist in a clarification of the fundamental structure of the theory, by pointing out the boundaries that the theory itself is supposed not to violate when satisfying a class of basic constraints. In fact, however, the history of the no-go theorems for quantum mechanics has been highly controversial: to cite a well-known instance, the original paper in which John S. Bell proved the theorem that bears his name contains a preliminary discussion on why the 'no-go theorems' existing at that time were far from showing what they were purported to show – namely that no hidden-variable completion of quantum mechanics was possible in principle. In spite of the controversies that such no-go industry generated, and of the intrinsic difficulty in justifying the several conditions required in the long chain of no-go theorems for quantum mechanics in the last decades, the attempt of establishing more and more stringent *negative* results continues to affect the imagination of many. In the present paper I will discuss some recent cases and question the significance of these results, with a two-fold strategy. First, I will consider three results concerning respectively local realism, quantum covariance and predictive power in quantum mechanics, and I will try to show how controversial the main conditions of the negative theorem turn out to be – something that appears to undermine the general relevance of these theorems. Second, I will try to discuss what I take to be a common feature of these theoretical enterprises, namely that of aiming at establishing negative results for quantum mechanics *in absence of a deeper understanding of the overall ontological content and structure of the theory*. I will argue that a sensible way toward such an understanding may be to cast in advance the problems in a clear and well-defined interpretational framework – which in my view means primarily to specify the ontology that quantum theory is supposed to be about – and *after* to wonder whether problems that seemed worth pursuing still are so in the framework. Although clearly this is not the *only* way, one can try to motivate it on the basis of a 'robust' view of what a foundational view of quantum mechanics should be.

The paper is organized as follows. The sections 2-4 will be devoted to a critical analysis of three recent no-go claims, concerning respectively the role of a 'realism' condition (section 2), the status of quantum covariance (section 3) and the issue of the predictive power of quantum states (section 4); finally, in section 5, I will attempt an assessment of the no-go strategy, also connecting it with the more general issue of what it really takes to 'interpret' quantum mechanics in a philosophically sensible way.

## 2 A No-Go Theorem about Quantum Realism: the Leggett Non-Local 'Realistic' Models

The first instance of recent no-go theorems that we will consider concerns 'realism' (the reason for using quotation marks for the term *realism* will be clear soon) and is based on a widespread interpretation of the Bell theorem, according which the theorem itself proves essentially that *local realism* is inconsistent with quantum theory. On the basis of this formulation of the Bell theorem, the question then arises whether there might be a *non-local* sort of realism that could be made consistent with quantum theory. Since the work of Leggett 2003, however, a class of non-local realistic theories has been introduced in order to show that not even inflating a suitable amount of non-locality into theories that simulate quantum theory (namely, that preserve quantum predictions) can allow them to be 'realistic': for it can be shown that an inequality can be derived within the class of non-local realistic theories that is violated by quantum theory, both theoretically (Leggett 2003) and experimentally (Gröblacher *et al.* 2007). As a consequence – here is the no-go theorem – 'realism' cannot be consistent with quantum theory, be it local (blocked by the violation of the Bell inequalities) or non-local (blocked by the violation of Leggett inequalities).

In strictly logical terms, the scope of a theorem depends on the generality of its assumptions so that, in order to assess to what extent 'realism' is really incompatible with quantum theory it is necessary to examine the very notion of realism that is adopted in establishing the above no-go result. If we do so, interestingly we discover two facts. First, the condition of 'realism' assumed in the Leggett framework is too strong. Second, the overall framework is motivated by the foundational point of view by an interpretation of the Bell theorem which is simply mistaken. Let us address the two points in turn.

## 2.1 Realism in the Leggett framework

In order to see why the 'realism' assumption is highly questionable in the Leggett approach, let us introduce the essential features of this framework, proposed for the first time in Leggett (2003). As a general premise, Leggett claims that

> Bell's celebrated theorem states that, in a situation like that considered by Einstein *et al.*, which involves the correlation of measurements on two spatially separated systems which have interacted in the past, no *local* hidden-variable theory (*or more generally, no objective local theory*) can predict experimental results identical to those given by standard quantum mechanics. (p. 1469, italics added)

Leggett proposes then introducing a class of *non-local* hidden-variable theories – namely a class of theories which, while retaining 'objectivity' (as will be seen later, it is an extra-strong assumption of realism), admits the possibility of non-local physical processes. The motivation for such a theoretical move is the following:

> In my view, the point of considering such theories is not so much that they are in themselves a particularly plausible picture of physical reality, but that by investigating their consequences one may attain a deeper insight into the nature of quantum-mechanical "weirdness" which Bell's theorem explores. *In particular I believe that the results of the present investigations provide quantitative backing for a point of view which I believe is by now certainly well accepted at the qualitative level, namely that the incompatibility of the predictions of objective local theories with those of quantum mechanics has relatively little to do with locality and much to do with objectivity.* (p. 1470, italics added)

The theories in the Leggett class are supposed to account for the results obtained in a general experimental framework, in which some polarization measurements are performed on pairs of photons emitted by atoms in a cascade process (Leggett (2003), p. 1471 ff). Since this framework encompasses, after the emission, a number of detection processes involving a pair of spatially separated detectors (let us call them **D**$_1$ and **D**$_2$), attention is focused as usual on correlations between the counts: clearly, the aim is to compare the predictions for a given function of such correlations as prescribed by quantum mechanics on the one hand and by (what Leggett assumes as) a general hidden-variable theory on the other.

The general conditions that the Leggett-type of theories are assumed to satisfy are the following (Leggett (2003), pp. 1473-4):

**L1**. Each pair of photons emitted in the cascade of a given single atom is characterized by a *unique* value of some set of hidden variables denoted by λ.

**L2**. In a given type of cascade process, the ensemble of pairs of emitted photon is determined by statistical distribution of the values of λ, characterized by a normalized distribution function ρ(λ). Such function is assumed to be *independent* of any parameter concerning polarizer settings (denoted in the sequel with **a** and **b**) and detection processes.

**L3**. If **A** and **B** denote respectively two variables that take the value +1 (−1) according to whether the detectors **D**$_1$ and **D**$_2$ register (do not register) the arrival of a photon, the value of **A** may depend not only on **a** and λ but also possibly on **b**, and similarly the value of **B** may depend not only on **b** and λ but also possibly on **a**.

Condition **L1** amounts to assuming *pre-existing* and *measurement-independent* properties for the pair of photons involved in the experiments, condition **L2** prevents the possibility of conspiratorial dependences between the source and any parameter involved in the spacetime regions where the polarizers and the detectors are located, whereas **L3** allows for possibly non-local influences of polarizer setting parameters on the outcomes[1]. Clearly this last condition, which according to the Leggett terminology characterizes the theories of the class as *crypto-nonlocal*, is where the theory is supposed to go beyond the class of theories that – *according to the Leggett and followers' approach*)[2] – are ruled out by the Bell theorem . Jointly, **L1**-**L3** imply the following expression for the correlation to be measured P(**a**, **b**)

$$P(\mathbf{a}, \mathbf{b}) = \int_\Lambda \mathbf{A}(\mathbf{a}, \mathbf{b}, \lambda)\, \mathbf{B}(\mathbf{b}, \mathbf{a}, \lambda)\, \rho(\lambda)\, d\lambda .$$

---

[1] It might be called 'Non-local determinism', since the actual outcomes are well determined by the pre-existing properties of the systems but possibly also in a non-local way.

In addition to **L1**-**L3**, it is assumed that the local averages ⟨**A**⟩ and ⟨**B**⟩ agree with the relevant quantum mechanical predictions, which appears to be a rather natural 'consistency' condition on the Leggett class of hidden-variable theories (Leggett (2003), pp. 1476-9, Gröblacher et al (2007), p. 872). The last step is then the statement of an 'incompatibility' result consisting in the derivation, within crypto-nonlocal realistic hidden-variable theories, of an inequality that is theoretically violated by the corresponding quantum mechanical expressions (Leggett (2003), sect. 3).

By *reductio*, the violation of the Leggett inequality testifies *essentially* against realism (in the Leggett formulation). In fact the Leggett theory assumes also another condition, namely crypto-nonlocality, so that in strictly logical terms the violation of the Leggett inequality implies the negation of the *conjunction*

<p style="text-align:center">realism AND crypto-nonlocality</p>

which, in turn implies

<p style="text-align:center">EITHER non-realism OR non-crypto-nonlocality [i.e. locality].</p>

But it is the first disjunct that must hold: we certainly cannot interpret the violation of the Leggett inequality as directed against crypto-nonlocality, since dropping this condition would make the theory a *local* theory, and we know from the Bell theorem that such a theory is inconsistent with quantum theory. Therefore, the real target is realism.

The problem is that, in the Leggett formulation, this sort of realism is absolutely too strong in order for this condition to be really significant and play the role that in principle the Leggett framework assigns to it. In fact the Leggett-sort of realism, by assuming that quantum systems have *pre-existing* and *measurement-independent* properties, turns out to be incompatible with quantum theory quite independently from any derivation of inequalities, so that the violation of certain inequalities by a non-local 'realistic' theory of the Leggett class can hardly tell against the consistency of any form of realism in the quantum realm. The Leggett realism turns out to be an unreasonable assumption of any allegedly 'objective' theory of quantum phenomena in a way that closely resembles all existing no-hidden variable theories proofs of the 60's (Gleason, Jauch-Piron, Kochen-Specker and an additional proof provided by Bell himself as a simplified version of the Kochen-Specker theorem): as shown already by Bell himself in the article that *preceded* the Bell theorem article, although it was published *after* it (Bell 1966), all these proofs required assumptions that it was not reasonable to require from any hypothetical completion of quantum theory[3], and it is no surprise that the Bohmian theory – namely the only serious existing 'hidden variable' theory – need not satisfy any of these assumptions (in fact, it violates them). As Bell remarked:

> It will be urged that these analyses [i.e. the above mentioned proofs] leave the real question untouched. In fact it will be seen that these demonstrations require from the hypothetical dispersion free states, not only that appropriate ensembles thereof should have all measurable properties of quantum mechanical states, *but certain other properties as well*. These additional demands appear reasonable when results of measurement are loosely identified with properties of isolated systems. They are seen to be quite unreasonable when one remembers with Bohr 'the impossibility of any sharp distinction between the behaviour of atomic objects and the interaction with the measuring instruments which serve to define the conditions under which the phenomena appear'. (Bell (1966), in Bell (2004), pp. 1-2, italics added)

---

[2] As a matter of fact, the Leggett-type of theories are 'realistic' hidden variable theories that are assumed to be non-local by accepting the condition usually called *outcome independence* but dropping the further condition called *parameter independence* (according to the Shimony revision of the terminology introduced in Jarrett 1984).. I wish also to stress that in presenting the Leggett framework I skip several technical details that, although deserving attention, are inessential to the present discussion.

If the Leggett realism were an *independent* assumption of any hidden variable theory, Gleason-Bell-Kochen & Specker would have already proved their incompatibility with quantum mechanics *needless of any locality requirement*. But, as Bell showed, there is little significance in testing against quantum theory a theory (be it local or non-local) that is supposed to satisfy a condition that we already know quantum mechanics cannot possibly and reasonably satisfy[4].

---

[3] Bell also mentions an especially restrictive assumption of the von Neumann theorem, an assumption which makes the von Neumann formulation much stronger with respect to the non-contextual formulations given by Gleason, Jauch-Piron, Bell and Kochen-Specker, and hence even less plausible.

[4] Bell recalls this point still in the opening page of his 1964 paper: "There have been attempts to show that even without such a separability or locality requirement no 'hidden variable' interpretation of quantum mechanics is possible. These attempts have been examined elsewhere and found wanting [Bell refers here to his 1966 article]" (Bell (1964), in Bell (2004), p. 14). More recently, the same charge has been clearly stated and motivated in Norsen (2007), pp. 317-8.

## 2.2 The interpretation of the Bell theorem in the Leggett framework

The second point is that the strategy leading to the no-go theorem against realism is based on a wrong interpretation of the Bell theorem in the first place, that undermines the whole project. The latter assumes in fact that the Bell theorem has among its assumptions *both* locality *and* realism, whereas it is easy to check that in the usual framework – namely, in the EPR setting with strict spin anticorrelation – the very existence of definite properties (call them 'hidden variables', 'pre-existent properties', 'objective properties', 'classical properties' or whatever) *is a consequence* of the locality assumption. Since in the EPR setting the distant spin outcomes turn out to be anticorrelated, if we require the theory to be local then it cannot be the case that the anticorrelation is explained by the measurement procedure on one side affecting the outcome at the other, far away side. *Therefore*, the only reasonable explanation of the distant spin outcomes being anticorrelated is that there are definite values for the spins already at the source: due to the logical structure of the argument, the only independent assumption is undoubtedly locality. But also in the more general EPR setting with *non*-strict spin anticorrelation, the so-called stochastic hidden-variable theories' framework (originally introduced in Bell 1971 and Clauser, Horne 1974), no independent 'realism' assumption plays any role although, once again, conventional wisdom tries its best to include it in the set of the Bell theorem's conditions. In the stochastic hidden-variable theories' framework (we will refer to the BCH framework, since this was originally introduced in Bell 1971 and Clauser, Horne 1974), a typical EPR joint system $S_1+S_2$ is prepared at a source, so that a 'completion' parameter $\lambda$ is associated with the single and joint detection counts. Suppose we denote by **a** and **b** respectively the setting parameters concerning two detectors, located at space-like separation and devised to register the arrival of $S_1$ and $S_2$ respectively. The model then is assumed to satisfy the following conditions:

**BCH1**    The parameter $\lambda$ is distributed according to a function $\rho(\lambda)$ that does *not* depend either on **a** or on **b**.

**BCH2**    The parameter $\lambda$ prescribes single and joint detection *probability*.

**BCH3**    *Locality* holds, namely the $\lambda$-induced probability for the measurement outcomes for $S_1$ and $S_2$ separately is such that (**i**) the detection probability for $S_1$ depends only on $\lambda$ and **a**, (**ii**) the detection probability for $S_2$ depends only on $\lambda$ and **b**, (**iii**) and the joint detection probability is simply the product of the detection probability for $S_1$ and the detection probability for $S_2$.

What Bell is interested to in this context is a joint probability distribution P(A, B | **a**, **b**), where each A and B represent given measurement outcomes and **a** and **b** stand respectively for the above mentioned setting parameters (with the obvious interpretation). No mention of what sort of systems are involved need be made, and once (rather) innocuous conditions on the probabilistic structure are assumed, it is easy to show the derivation of an inequality that turns out to be violated by the corresponding quantum correlations. According to one of the recent anti-realistic claims, however, among the assumptions of the stochastic version of the Bell theorem there is still realism, defined as follows:

"*Realism*. To put it short: results of unperformed measurements have certain, unknown but fixed, values. In Bell wording this is equivalent to the hypothesis of the existence of hidden variables" (Žukowski 2005, p. 569).

But, again, such an assumption need not be required. It is obviously true that in the stochastic framework locality does not imply by itself the existence of definite spin properties, because the stochastic framework does not encompass *strict* anticorrelation. Nevertheless, assuming the existence of such properties is unnecessary: the core of the argument lies simply in stating what preventing any action-at-a-distance amounts to, *whatever the factors determining A and B might be*. A realism-flavoured additional assumption, according to which there are some pre-existing properties in the common past of the relevant events at A and B that enhance the correlation, is simply irrelevant: should such an assumption be adopted, it would be obviously sufficient for the existence of local factors, but it would such a strong requirement as to make virtually empty the class of 'serious' local theories that might be put to test in a stochastic framework. In other words, it is true that the assumption of pre-existing properties for the two systems at the source might well imply locality, but the assumption that only local operations and influences can contribute to fix the single detection probabilities does *not* require the assumption of pre-existing properties[5].

Summing up: if the whole point of the Bell argument (also in the stochastic case) is then in fact to show that the correlations between the results A and B are not locally explicable, no matter what the relation is between A and B on one side and some allegedly 'objective' or 'pre-existing' properties corresponding to them on the other, we can safely say that also in the more general (no strict correlation) case, *there is no 'realism' at stake*.

---

[5] For an exceptionally clear statement see Bell 1981, in Bell 2004, p. 150.

It is interesting to remark that the 'local-realistic' reading of the Bell theorem and its meaning, however, continues to be widespread and can be found also in different formulations. In a recent review paper on the Bell inequalities and their relevance to quantum information theory, Brukner and Žukowski depict the situation in terms of the following experimental framework (Brukner and Žukowski 2010). At two different stations of a typical EPR-like arrangement, stations that are supposed to be sufficiently far away from each other and that we will call *A* and *B*, Alice and Bob are endowed each with a display, on which they observe sequences of +1 and −1 appearing. With respect to a selected reference frame, the numbers appear simultaneously and are caused to appear on Alice's and Bob's displays by the activation of a 'source', located in the middle between the two station. Moreover, the two stations have each two possible 'settings': if we denote with $m = 1, 2$ the possible settings at *A*, and with $n = 1, 2$ the possible settings at *B*, a random, local procedure is supposed to take place at each station in order to select a specific setting at each station. Now, according to Brukner and Žukowski, it is reasonable to account for the above situation by a *local-realistic* model, i.e. a model that satisfies the assumptions of **REALISM**, **LOCALITY** and **FREE WILL**, namely:

**REALISM** – Given the eight variables $A_{m,n}$, $B_{n,m}$ with $n$, $m$ ranging over $\{1,2\}$, the expression

$$A_{m,n} = \pm 1$$

is meant to indicate that the value at *A* is ± 1 provided that the setting at *A* is *m* and the setting at
*B* is *n*. This is equivalent to the assumption that a joint probability distribution

$$P(A_{1,1}, A_{1,2}, A_{2,1}, A_{2,2}; B_{1,1}, B_{1,2}, B_{2,1}, B_{2,2})$$

always exists.

**LOCALITY** – The appearance of a given value on the display at Alice's (Bob's) station in no way depends on what happened at Bob's (Alice's) station. The expression 'what happened' includes both the selection of a given setting and the appearance of a specific value.

**FREE WILL** – The selection of a local setting at a given station (be it *A* or *B*) in no way depends on the source.

On the basis of these assumption, Brukner and Žukowski show that a CHSH-type inequality can be easily derived (Brukner and Žukowski 2010, eq. (23)).

What does the point seem to be about realism, then? The point seems to be the assumption that realism is equivalent to the *existence* of the joint probability distribution

$$P(A_{1,1}, A_{1,2}, A_{2,1}, A_{2,2}; B_{1,1}, B_{1,2}, B_{2,1}, B_{2,2}).$$

But one thing is to define what realism amounts to, and quite another one to assume that the definition is *actually satisfied*: I can well define what a winged horse is supposed to be, without being able to prove that such a thing exists in the world! As a matter of fact, in the above model the characterization of **REALISM** as the existence of a suitable joint probability distribution does not imply *by itself* that such a distribution exists: it is exactly **LOCALITY** that imposes on the form of the distribution the very constraint we need in order to be sure that the desired joint probability distribution actually exists. For let us assume that the theory is local. Then $A_{m,n} = A_m$ and $B_{n,m} = B_n$,

from which

$$P(A_{1,1}, A_{1,2}, A_{2,1}, A_{2,2}; B_{1,1}, B_{1,2}, B_{2,1}, B_{2,2}) = P(A_1, A_2, B_1, B_2).$$

Due to **LOCALITY**, therefore, we are sure that a joint probability distribution like $P(A_1, A_2, B_1, B_2)$ certainly exists, since we can always set

$$P(A_1, A_2, B_1, B_2) = P(A_1)\,P(A_2)\,P(B_1)\,P(B_2),$$

where the distribution $P(A_1)\,P(A_2)\,P(B_1)\,P(B_2)$ is trivially compatible with the distributions

$$P(A_1\,\&\,B_1),\ P(A_1\,\&\,B_2),\ P(A_2\,\&\,B_1),\ P(A_2\,\&\,B_2)$$

as marginals, since $P(A_n\,\&\,B_m) = P(A_m)\,P(B_n)$, with $n, m = 1,2$. Also in this framework, that is, realism is justified by locality which turns out then to be the real culprit.

The overall philosophical lesson that we can learn from the above discussion is twofold. First, it seems pointless to launch a no-go anathema against realism by 'inflating' an *a priori* notion of realism (in terms of pre-existing properties) into quantum theory, only to discover that quantum theory itself cannot possibly host that notion no matter whether it is local or not! Second, if this sort of realism is inconsistent with quantum theory from scratch, it makes little sense to require it from any hypothetical local extension of quantum theory, since the latter might be ruled out due to the unreasonable 'realism' assumption. On these grounds the derivation of the Leggett inequalities, then, can hardly have the status of a serious 'no-go theorem', so that we can safely claim that the question is open whether realism can still play a conceptual role in the philosophical foundations of quantum theory.

## 3  A No-Go Theorem about Quantum Covariance: The Gisin Models

A second case study is provided by a recent no-go claim according to which it is impossible in principle to construct a deterministic nonlocal hidden variable extension of quantum theory that satisfies a particular covariance requirement (Gisin 2011). According to (a correct reading of) the Bell theorem, we know from that no (deterministic or stochastic) extension of quantum theory – *be it a hidden variable theory or not* [6] – can be local; we may still ask, however, whether there exists some sort of deterministic *nonlocal* (hidden variable) extension with different properties, in particular with respect to relativistic covariance. In his recent paper, Gisin shows that no such extension is possible if we require that the extended theory be covariant, namely no such extension can account for quantum correlations in the sense of relativistic time-order-invariant predictions (namely under the change of time ordering).

The framework is a typical Bell experimental setting, with the emission of pairs of spin-1/2 particles prepared at the source in the spin singlet state $\psi^{(-)}$. In this ideal setting the source state of the joint system prescribes a strict anticorrelation between the measurement outcomes in the two wings of the experimental setting, whereas the measurement outcomes were supposed to be associated with spacetime regions that are space-like separated. According to the usual terminology, we will call Alice and Bob the two parties involved in the two distant kinds of measurements. Since the two measurement regions are space-like separated, there will be frames $F$ in which Alice precedes Bob (in the $F$-time-ordering) in choosing her setting parameter **a** and frames $F'$ in which such time ordering is reversed, so that Bob is first in choosing his setting parameter **b**. In addition to this, the Gisin model assumes a so-called 'nonlocal' variable $\lambda$, which is supposed to determine the measurement result.

Suppose now that Alice chooses *first* her setting **a**: her measurement result $\alpha$ can be expressed as a function of the initial state of the composite system $\psi^{(-)}$, of the setting parameter **a** and of the nonlocal variable $\lambda$, namely

(1) $\qquad\qquad\qquad \alpha = F_{AB}(\psi^{(-)}, \mathbf{a}, \lambda)$

whereas the measurement result for Bob, who chooses his setting **b** *after*, can be expressed as a function of the initial state of the composite system $\psi^{(-)}$, of the setting parameter **b**, of the nonlocal variable $\lambda$ *but also of the setting parameter* **a**, namely

---

[6] It has still to be stressed how much confusion in the debates concerning the meaning of the Bell theorem derives from insisting on the requirement that the supposed local extension of quantum theory (an extension that the Bell theorem proves to be impossible) has to be also a *hidden variable* theory.

(2) $$\beta = S_{AB}(\psi^{(-)}, \mathbf{a}, \mathbf{b}, \lambda).$$

(Clearly the letters 'F' and 'S' denote who was the first and who the second in choosing the setting.) Gisin stresses that the model is assumed to be 'non-local' exactly because Bob's result might depend also on **a**, in addition to **b**: "this is the sense in which the variable $\lambda$ *together with the functions* $F_{AB}$ *and* $S_{AB}$ *form a nonlocal model*" (our emphasis). If we consider the reverse time order, in which it is Bob who chooses first, we will have

(3) $$\beta = F_{BA}(\psi^{(-)}, \mathbf{b}, \lambda)$$

and

(4) $$\alpha = S_{BA}(\psi^{(-)}, \mathbf{a}, \mathbf{b}, \lambda),$$

where the functions $F_{BA}$ and $S_{BA}$ need not coincide with the functions $F_{AB}$ and $S_{AB}$, respectively, due to the different time order.

Now a covariance condition is required from such a nonlocal model, namely the condition according to which Alice's result, being a scalar, should be independent of the reference frame. Hence we obtain for $\alpha$ and $\beta$

(5) $$\alpha = F_{AB}(\psi^{(-)}, \mathbf{a}, \lambda) = S_{BA}(\psi^{(-)}, \mathbf{a}, \mathbf{b}, \lambda)$$

(6) $$\beta = F_{BA}(\psi^{(-)}, \mathbf{b}, \lambda) = S_{AB}(\psi^{(-)}, \mathbf{a}, \mathbf{b}, \lambda)$$

from which we derive that (5) makes the 'Bob-first' model a *local* model, whereas (6) makes the 'Alice-first' model a *local* model. In turn, this means that any such covariant model is equivalent to a local model in the sense of Bell, a consequence according to which no covariant hidden variable extension of quantum theory is possible since it is committed to the Bell inequality which is violated by quantum theory.

What is the awkward feature of this otherwise simple model? It seems then that the 'non-local' character of $\lambda$ shows up only when the setting **a** becomes relevant for Bob and the setting **b** becomes relevant for Alice. In other words, $\lambda$ is nonlocal only for Bob in the form of [$\lambda$ + **a**] when Bob's measurement comes second, and nonlocal only for Alice in the form of [$\lambda$ + **b**] when Alice's measurement comes second. In this way, the nonlocality of the model would manifest only for a single party at a time, so that – as a matter of fact – the model is *local* just for the party that comes first. But if by Covariance $S_{BA}(\psi^{(-)}, \mathbf{a}, \mathbf{b}, \lambda)$ is equated to something like $F_{AB}(\psi^{(-)}, \mathbf{a}, \lambda)$ – which was *local* at the beginning! – then it is not covariance that must be blamed for the model becoming local, but rather the fact that λ is just named 'nonlocal' but works from the beginning as a *local* variable.

Let us even suppose that this awkward feature can be justified by the idea that a measurement that comes first cannot depend on a choice that has yet to be made: but what happens in the situation – allowed by the space-like separation of the two measurements – in which we select a frame of reference in which the two measurements are *simultaneous*? Is the model non-local or not?

Let us denote by

$$\alpha = \text{SIM}_{AB}(\psi^{(-)}, \mathbf{?}, \lambda)$$

the result of Alice's measurement when Bob's measurement is performed simultaneously: what should the question mark be replaced with? If we write

$$\alpha = \text{SIM}_{AB}(\psi^{(-)}, \mathbf{a}, \lambda)$$

then the model is already assumed to be local and there is no point to require it to be covariant, since it cannot work anyway. But if we write instead

$$\alpha = \text{SIM}_{AB}(\psi^{(-)}, \mathbf{a}, \mathbf{b}, \lambda)$$

then we should have, by Covariance, that

$$\alpha = \text{SIM}_{AB}(\psi^{(-)}, \mathbf{a}, \mathbf{b}, \lambda) = F_{AB}(\psi^{(-)}, \mathbf{a}, \lambda) = S_{BA}(\psi^{(-)}, \mathbf{a}, \mathbf{b}, \lambda)$$

The model, then, seems local or non-local according to the frame of reference: in turn, this makes unclear the idea that Covariance turns a *clearly* nonlocal model into a *clearly* local model.

# 4  A No-Go Theorem about the Predictive Power of Quantum Theory: the Colbeck-Renner 'Free' Models

In a recent paper, Colbeck and Renner establish a further no-go result, concerning the predictive power of quantum theory (Colbeck, Renner 2011). As is well known, quantum theory is an indeterministic theory in a precise sense: given a quantum system S in the state ρ and a physical quantity Q measurable on S, the state ρ does not determine in general the outcome of the measurement but provides only the probability of recording one of the possible outcomes with a certain probability. Since there is no available proof that this represents the maximal possible amount of information that can be gathered from a quantum system[7], Colbeck and Renner raise the question whether there exist extensions of quantum theory (not necessarily taking the form of hidden variable theories) that might improve the extent of the information concerning the outcome in a measurement process. Answering in the negative, Colbeck and Renner provide a further instance of a no-go theorem in the foundations of quantum theory under a pair of general assumptions that they take to be very natural.

The first assumption (called FR), whose status has been widely discussed in the past[8], concerns the possibility of freely choosing the measurement settings. In order to be correctly formulated, FR depends on the notion of *spacetime random variable* (denoted by SV), namely a random variable together with a set of coordinates $(t, r_1, r_2, r_3) \in \Re^4$. Intuitively, a SV at a given spacetime point $(t, r_1, r_2, r_3)$ is to be interpreted simply as a value that is accessible at $(t, r_1, r_2, r_3)$, so that a measurement process in this context is just a process that takes an input $A$ to an output $X$, where both $A$ and $X$ are SV's. According to Colbeck and Renner, the assumption FR is nothing but the assumption that we can always select a SV $A$ as the input of a measurement process such that all the other SV lying outside the future lightcone turn out to have no correlation with $A$. The second (uncontroversial) assumption is the validity of QM.

On the basis of these assumption[9], Colbeck and Renner show the following. If we denote with $A$ a measurement setting and with $X$ the measurement outcome, we know that quantum theory describes both $A$ and $X$ in such a way as to generate a probability distribution $\mathbf{P}_{A/X}$ of the possible outcomes $X$ for any given $A$. Should this distribution be not informationally maximal, there should be some further information $\Xi$ that, 'added' somehow to the information encoded in the distribution, is supposed to improve the precision in the prediction of the outcome. What Colbeck and Renner prove is essentially that $\Xi$ is irrelevant: the distribution $\mathbf{P}_{A+\Xi/X}$, namely the distribution of $X$ given $A$ *plus* the additional information $\Xi$, is equal to the former distribution $\mathbf{P}_{A/X}$, so that $\Xi$ does not seem contribute to improve the predictive power associated to the standard quantum distribution (Colbeck, Renner 2011, p. 3).

The framework in which this result is obtained is intentionally formulated in a very general way for at least two reasons: first, in order to encompass the most general possible analysis of a measurement process in quantum theory and, second, to envisage the most general possible

---

[7] In fact, that quantum theory is 'complete' in this sense is a *postulate* of the theory and not a theorem that can be proved within the theory.
[8] At least since the Bell 1977 paper "Free variables and local causality" (in Bell 2004, pp. 100-104).
[9] We put technical details aside, but the interested reader may consult the *Supplementary Information* attached to the main paper, which is presented in a rather qualitative form.

framework "within which any arbitrary extra information provided by an extension of the current theory can be considered" (Colbeck, Renner 2011, p. 2). In this very general framework, however, the hypothetical additional information Ξ is supposed to satisfy a pair of assumptions that are innocuous only at first sight. Although Colbeck and Renner state explicitly concerning Ξ that "we do not assume that it is encoded in a classical system but, instead, we characterize it by how it behaves when observed" (Colbeck, Renner 2011, p. 3), they require that such informational entity is such that

(i)     we can have access to it at any time;
(ii)    it is static, namely that its behavior does not depend on where or when it is observed.

In order to evaluate the relevance of the Colbeck-Renner result, we may again wonder whether the alleged generality of their model is really a virtue. The opposite seems to be the case and we can realize it by taking into account what appears to be the only serious candidate for being an informationally finer version of QM, namely Bohmian mechanics[10]. The latter starts from the assumption that the ordinary quantum wave function need not be complete – in itself a perfectly plausible and consistent assumption to make – and adds a further element in order to complement the informational role of the wave function itself, namely the particle positions. What matters here is that Bohmian mechanics locates this addition within an ontologically robust and non-ambiguous image of the quantum world and, above all, does it *without forcing on the extra-bit of information either that it be always accessible or static*. In the framework of Bohmian mechanics, we have a fairly straightforward explanation of how the assumption of extra-elements that are potentially relevant in informational terms can coexist with the ordinary QM statistics for any experiment that we can perform, namely just the circumstance that the extra elements *are only partially accessible*. For these reasons, disregarding the actual structure of a theory which presents itself as an alternative to ordinary QM, but able to recover all the statistical content of the latter, is actually a vice rather than a virtue.

Moreover, this attitude leads Colbeck and Renner to attribute to Bohmian mechanics a wrong position with respect to the assumption FR, since they claim that "in the context of de Broglie-Bohm theory, the presence of non-local hidden variables contradicts assumption FR" (Colbeck, Renner 2011, p. 3). The 'freedom' in the choice of measurement settings is relevant at the level of statistics: a possible way in which the choice might be not totally free would show up if some event lying outside the lightcone could affect the process so as to alter the measurement statistics. But Bohmian mechanics is bound for construction to agree on QM measurement statistics and need not violate FR, inasmuch as the latter constrains what occurs at the statistical level. It is true that Bohmian mechanics is non-local, in that it encompasses influences travelling outside the lightcone: but for the reason sketched above, such influences get washed out of the statistical level, which is the level that matters concerning the measurement process. So, contrary to what Colbeck and Renner claim, the reason why Bohmian mechanics is not covered by their result is rather that it fails to satisfy the accessibility condition on $\Xi$ [11].

---

[10] We wish to stress that we refer to Bohmian mechanics for exemplifying and explanatory reasons, without claiming any apriori superiority for it with respect to competing views on the foundations of quantum mechanics. True, we feel that the Bohmian framework is able to address foundational issues with a clarity and crispness that are often rare in current debates but this has to do with a personal – as such, highly debatable – philosophical taste.

## 5  The No-Go Strategy: what's basically wrong with it?

As we have seen, the supporters of the no-go strategy usually assume the extreme generality of their models as a virtue: we do not need to enter in too many details – they would probably argue – in order to show that we cannot extend orthodox quantum theory into a theory of quantum phenomena that preserves properties that some might like to retain – be they locality, realism, covariance or predictive power. The main point of the present paper is that that generality, far from being a virtue, is quite the opposite. It is *exactly* their being abstract and detached from the *actual* alternatives to quantum theory that deprives the models proposed by the no-go strategies of a deeper significance. Not only all these results take ordinary quantum theory itself at face value, so that the extending theories are supposed to inherit all the vagueness implicit in the orthodox treatment of the basic notions of ordinary quantum mechanics (clearly, the most urgent vagueness being the meaning of the wave function). The no-go results that we have discussed above, although addressing different issues, display an underlying common feature, that of avoiding any reference to a more detailed conceptual structure of the hypothetical theory that should extend or replace ordinary quantum theory.

---

[11] A last critical remark. The authors claim that FR involves the lightcone structure of the relativity theory just to provide a meaning to the idea of spacetime random variable, and "does not involve any assumptions about relativity theory" (p. 2). First, assuming the lightcone structure *is* basically assuming relativity theory. Second, taking seriously the fact that the spacetime which is the arena of quantum measurements is at least a special-relativistic spacetime is essential in several respects: otherwise awkward properties would hold for QM, first of all signalling across distant regions since in a non-relativistic spacetime no limit to the travel of information can be postulated.

In the case of the Leggett-type of theories, an extra-strong kind of 'realism' is simply imposed from outside, without any regard to the sort of much more sophisticated kind of realism can happily survive in different, more hospitable frameworks (such as Bohmian mechanics or GRW theories). In the case of the Gisin model, proving that no deterministic non-local hidden variable extensions of QM can be covariant conveys no useful information on the *inner* structure and character of any of these extensions. Once again the case of Bohmian mechanics can help. This theory is a non-local deterministic hidden variable theory which is endowed with a structure that *does* justify its failure in satisfying covariance: as soon as the theory is required to describe a system of *n* particles, the guiding equation refers to a unique, absolute time for all *n* particles, a circumstance that makes the description a non-Lorentz invariant one. This justification makes the non-covariance of Bohmian mechanics remarkably more relevant and instructive by a foundational point of view, since its non-Lorentz invariance is deeply connected with the overall structure of the theory itself. As to the Colbeck-Renner result, the situation is even more striking, since a focus on the issue of finding a theory that might be interpreted as 'informationally finer' compared to QM should have led naturally toward a fair assessment of such a theory as Bohmian mechanics but, ironically enough, that theory just happens *not* to satisfy at least one of the main conditions under which the Colbeck-Renner is established.

A more general remark concerning the no-go strategy is in order, a remark connected with one of the hottest issues under debate since the birth of the ordinary formulation of QM in the Thirties, namely the measurement problem. Unless one disagrees on the idea that there is just *one* physical world – which is the (logically consistent) option pursued the many-worlds interpretation – the measurement problem shows that the coupling between system and apparatus that ordinary QM prescribes gives rise to a physical situation that simply does not match with what we observe in the laboratory, unless we add some rules of thumb like the reduction of the wave function. The measurement problem is only the clearest sign of a fundamental fact: if we require from such a theory as QM something more than just predicting statistical patterns – something like answering the question: what would be the world like, should QM be (approximately) true? – then we have learned from decades of foundational debates that QM must be *supplemented* with additional structure, rather than deprived of it. Clearly, it is very hard to agree on many deep questions that this fundamental fact raises: What should this additional structure be like and how should it be interpreted?

Should it necessarily entail new experiments and/or new predictions or a better understanding of the theory's structure should be considered already a progress? Nevertheless, in view of the above discussed problems with the no-go philosophy, I argue that the only way toward such an understanding may be to cast in advance the problems in a clear and well-defined interpretational framework – which in my view means primarily to specify the ontology that quantum theory is supposed to be about – and *after* to wonder whether problems that seemed worth pursuing are still so in the framework. True, the question concerning the meaning of 'interpretational framework' admits different answers but I think we can still subscribe to the position of Richard Healey who, yet in his *The Philosophy of Quantum Mechanics. An Interactive Interpretation* (1989), wrote:

A satisfactory interpretation of quantum mechanics would involve several things. It would provide a way of understanding the central notions of the theory which permits a clear and exact statement of its key principles. It would include a demonstration that, with this understanding, quantum mechanics is a consistent, empirically adequate and explanatorily powerful theory. And it would give a convincing and natural resolution of the "paradoxes". I should like to add a further constraint: *that a satisfactory interpretation of quantum mechanics should make it clear what the world would be like if quantum mechanics is true*. (Healey 1989, p. 6)

The search for negative results of a general sort seems to hide the implicit tendency to avoid or postpone the *really* hard job: the attempt to interpret quantum mechanics according to a foundationally robust sense of 'interpretation', namely the attempt to make sense of it within a scientific image of the world in which we strive to understand what nature is and not what our theorizing is forced to be silent on[12].

---

[12] Two further examples worth mentioning in the context of the present paper are the Conway and Specker's *Free Will Theorem* (Conway, Kochen 2006), criticized among others by Tumulka 2007 and Goldstein, Tausk, Tumulka, Zanghì 2010, and the Hardy's *Quantum ontological excess baggage theorem* (Hardy 2004), criticized by Yong 2010.